\documentclass[10pt,conference,letter]{IEEEtran}

\usepackage{times}

\usepackage{amssymb}
\usepackage{amsthm}
\usepackage{amsxtra}
\usepackage{amsmath}
\usepackage{array}
\usepackage{algpseudocode}
\usepackage{algorithm}

\usepackage{verbatim}
\usepackage{epsfig}
\usepackage{setspace}
\usepackage{caption}
\usepackage{subcaption}
\usepackage{breqn}

\allowdisplaybreaks

\newtheorem{theorem}{Theorem}
\newtheorem{definition}{Definition}
\newtheorem{lemma}[theorem]{Lemma}

\newtheorem{example}{Example}[section]
\newtheorem*{conjecture*}{Conjecture}

\newtheorem*{remark*}{Remarks}

\def\CI{\text{CI}}

\def\CO{\text{CO}}
\def\SK{\text{SK}}

\def\cP{\mathcal{P}}
\def\cS{\mathcal{S}}
\def\cE{\mathcal{E}}
\def\cF{\mathcal{F}}
\def\cG{\mathcal{G}}
\def\cM{\mathcal{M}}
\def\cV{\mathcal{V}}

\def\BJ{\textsf{J}}
\def\BL{\textsf{L}}

\def\BF{\textsf{F}}
\def\bw{\mathbf{w}}
\def\tbw{\tilde{\bw}}
\def\tX{\tilde{X}}
\def\cH{\mathcal{H}}
\def\Ixm{I(X_{\mathcal{M}})}
\def\Ipxm{I_{\mathcal{P}}(X_{\mathcal{M}})}

\def\cR{\mathcal{R}}

\def\cC{\mathcal{C}}

\def\Ixml{I(X_{\cM}^n|\textsf{L})}
\def\gub{\Gamma}
\def\gs{\Gamma^*}

\def\bx{\mathbf{x}}

\def\N{\mathbb{N}}

\def\beq{\begin{equation}}
\def\eeq{\end{equation}}

\title{Bounds on the Communication Rate\\ Needed to Achieve SK Capacity \\ in the Hypergraphical Source Model}

\author{
\IEEEauthorblockN{Manuj Mukherjee$^\dag$} \and \IEEEauthorblockN{Chung Chan$^\ddag$} \and \IEEEauthorblockN{Navin Kashyap$^\dag$} \and \IEEEauthorblockN{Qiaoqiao Zhou$^\ddag$}
}

\begin{document}

\maketitle

\renewcommand{\thefootnote}{}
\footnotetext{
\noindent $^\dag$M.\ Mukherjee and N.\ Kashyap are with the 
Department of Electrical Communication Engineering, 
Indian Institute of Science, Bangalore. Email: \{manuj,nkashyap\}@ece.iisc.ernet.in.

\noindent $^\ddag$C.\ Chan and Q.\ Zhou are with the Institute of Network Coding at the Chinese University of Hong Kong. Email: cchan@inc.cuhk.edu.hk, qiaoqiaobupt@gmail.com.

\smallskip
}

\begin{abstract}
In the multiterminal source model of Csisz{\'a}r and Narayan, the communication complexity, $R_{\SK}$, for secret key (SK) generation is the minimum rate of communication required to achieve SK capacity. An obvious upper bound to $R_{\SK}$ is given by $R_{\CO}$, which is the minimum rate of communication required for \emph{omniscience}. In this paper we derive a better upper bound to $R_{\SK}$ for the hypergraphical source model, which is a special instance of the multiterminal source model. The upper bound is based on the idea of fractional removal of hyperedges. It is further shown that this upper bound can be computed in polynomial time. We conjecture that our upper bound is tight. For the special case of a graphical source model, we also give an explicit lower bound on $R_{\SK}$. This bound, however, is not tight, as demonstrated by a counterexample.
\end{abstract}

\renewcommand{\thefootnote}{\arabic{footnote}}

\section{Introduction}\label{sec:intro}

The problem of secret key (SK) generation for multiple terminals observing i.i.d.\ sequences of correlated random variables was first studied by Csisz{\'a}r and Narayan in \cite{CN04}. The terminals are allowed to communicate \emph{interactively} over a public noiseless channel. After the communication the terminals must agree upon an SK, secured from any eavesdropper having access to the public channel. The \emph{SK capacity}, i.e., the maximum rate of secret key that can be generated was derived in \cite{CN04}. A quantity of interest in the SK generation problem is the \emph{communication complexity}\footnote{Our use of ``communication complexity'' differs from the use prevalent in the theoretical computer science literature where, following \cite{Yao79}, it refers to the total amount of communication, in bits, required to perform some distributed computation.}, $R_{\SK}$, which is the minimum rate of communication required to generate an SK of maximum rate. 

Tyagi in \cite[Theorem 3]{Tyagi13} has given a complete characterization of $R_{\SK}$ for the case of two terminals. Tyagi's arguments have been extended by \cite[Theorem 2]{MNY} to give a lower bound on $R_{\SK}$ for the general multiterminal setting. This lower bound was computed and was shown to be tight for a special class of sources in \cite[Theorem 6]{MNY}. However, computing this lower bound for a general multiterminal source remains an open problem. Also, \cite{MNY} did not provide any discussion on the tightness of this lower bound. Hence, it is useful to derive upper bounds on $R_{\SK}$. The SK generation protocol in \cite{CN04} goes through \emph{omniscience}, i.e.,  all the terminals recovering the entire information of all the other terminals. Thus, the \emph{minimum rate of communication for omniscience}, $R_{\CO}$, is a valid upper bound on $R_{\SK}$. 

In this paper, we consider a special case of the multiterminal source model, namely the \emph{hypergraphical source model} studied previously in \cite{CZ10} and \cite{MNY}. The hypergraphical source model is inspired by the coded co-operative data exchange (CCDE) problem introduced in \cite{ESS10}, and has been studied in the context of the ``one-shot" SK generation problem, in \cite{CW14}--\cite{CH14}. One can also view the hypergraphical source model as a generalization of the pairwise independent network (PIN) model of \cite{NYBNR10} and \cite{NN10}. The main contribution of this paper is an upper bound on $R_{\SK}$ for the hypergraphical source model. The proof of this upper bound is based on the idea of \emph{decremental SK agreement} studied in \cite{Chan16}. The idea is to keep on removing ``randomness" from the hyperedges as long as the SK capacity does not decrease, and then use the $R_{\CO}$ of the resulting hypergraph as an upper bound on $R_{\SK}$ of the original hypergraph. We further show that the upper bound on $R_{\SK}$ thus derived is at least as good as $R_{\CO}$. Computation of this upper bound requires the solution of a linear program, whose separation oracle performs submodular function minimization. As a result the bound is computable in polynomial time. In fact, for the special case when the underlying hypergraph of the source model is a graph, the upper bound reduces to a simple expression. We believe that the upper bound is actually tight. Unfortunately, we do not have a proof of this yet, and therefore we state it as a conjecture. We also give a simple expression for the lower bound on $R_{\SK}$ derived in \cite[Theorem~2]{MNY}, for sources defined on graphs. Using this expression we are able to construct an example and show that the lower bound in \cite{MNY} is not tight in general. 

We would like to compare and contrast our work with those of Courtade et al. in \cite{CHISIT14} and \cite{CH14}. Courtade et al. consider a ``one shot" model with each terminal observing only one instance of a random variable. They restrict the communication to linear functions of the source randomness. In \cite[Theorem 11]{CH14}, they evaluate the minimum number of bits of communication required to generate a fixed number of bits of SK. It is also shown in \cite[Theorem 4]{CH14} that there exist sources where non-linear communication can strictly outperform any linear communication, in terms of the number of bits of communication. On the other hand, our major focus is the asymptotic model involving i.i.d. sequences of correlated random variables at each terminal. We also do not impose any linearity restriction on the communication. However, we consider communication complexity for generating SKs of maximum rate only, contrary to the arbitrary number of bits of SK considered by Courtade et al. It should be mentioned here that the proofs of Courtade et al. proceed by finding ``inherently connected subhypergraphs" obtained by completely removing certain hyperedges, as opposed to the ``fractional removal" of hyperedges in our proofs. The difference is due of the fact that Courtade et al. consider a one-shot model, whereas we look into an asymptotic scenario. Therefore, restricting ourselves to complete removal of hyperedges only will lead to weaker upper bounds, as demonstrated in Example~\ref{ex:square} appearing later in the paper.

The paper is organized as follows. The basic definitions and concepts are introduced in Section~\ref{sec:prelims}. Section~\ref{sec:ub} presents our main result, an upper bound to $R_{\SK}$. In Section~\ref{sec:lb}, we evaluate the lower bound to $R_{\SK}$ stated in \cite[Theorem 2]{MNY}, for the special case of graphical source models. The paper concludes with Section~\ref{sec:conc}.

\section{Preliminaries}\label{sec:prelims}

In this section we will introduce the major concepts and definitions used in this paper. Throughout, we use $\N$ to denote the set of positive integers. A \emph{weighted hypergraph} is defined by the pair $\cH=(\cM,\bw)$ with $\cM=\{1,2,\ldots,m\}$ denoting the set of vertices and $\bw: 2^\cM\to\mathbb{R}^+\cup\{0\}$ being the weight function on the subsets of the vertices. We will often view $\bw$ as a vector whose coordinates are indexed by $e\subseteq\cM$. The set of \emph{hyperedges} is obtained from the weight function as the set $\cE\triangleq\{e\in2^{\cM}: \bw(e)>0\}$ of subsets of $\cM$ with non-zero weights.

We say that a random vector $(X_i:i\in \cM)$ is a \emph{hypergraphical source} defined on the weighted hypergraph $\cH$ if we can write $X_i=(\xi_e:e\in \cE,i\in e)$ for some random variables $\xi_e$'s such that $H(\xi_e)=\bw(e)$ and $\xi_e$'s are mutually independent across $e\in \cE$. Whenever $\cE$ consists only of subsets $e$ of size 2, we refer to the source as a \emph{graphical source}, and refer to the hyperedges as \emph{edges}.  

For the hypergraphical source, $\cM$ denotes a set of terminals. Each terminal $i \in \mathcal{M}$ observes $n$ i.i.d.\ repetitions of a random variable $X_i$. The $n$ i.i.d.\ copies of the random variable are denoted by $X_i^n=(\xi_e^n: e\in\cE, i\in e)$.\footnote{Each i.i.d. sequence of random variables $\xi_e^n$, $e\in\mathcal{E}$, should be thought of as an SK initially shared among the terminals in $e$.} 
For any subset $A\subseteq \mathcal{M}$, $X_A$ and $X_A^n$ denote the collections of random variables $(X_i:i \in A)$ and $(X_i^n: i \in A)$, respectively. It is easy to check that $H(X_A)=\sum_{e\in\cE:e\cap A\neq\emptyset}\bw(e)$ and $H(X_A|X_{A^c})=\sum_{e\in\cE:e\subseteq A}\bw(e)$. 

We point out here that the hypergraphical source model is a special case of the multiterminal source model of \cite{CN04}, which is defined for an arbitrary joint distribution of $X_{\cM}$ over a finite support size. Note that the hypergraphical models studied in \cite{CZ10} and \cite{MNY} are but a special case of the model studied here, obtained by restricting $\bw$ to be integer-valued. If we further restrict $\bw(e)$ to take non-zero values only for subsets $e\subseteq\cM$ of size 2, we obtain the pairwise independent network (PIN) model of \cite{NN10}.

The terminals communicate through a noiseless public channel, any communication sent through which is accessible to all terminals and to potential eavesdroppers as well.
An \emph{interactive communication} is a communication $\textbf{f}=(f_1,f_2,\cdots,f_r)$ with finitely many transmissions $f_j$, in which any transmission sent by the $i$th terminal is a deterministic function of $X_i^n$ and all the previous communication, i.e., if terminal $i$ transmits $f_j$, then $f_j$ is a function only of $X_i^n$ and $f_1,\ldots,f_{j-1}$.   We denote the random variable associated with $\textbf{f}$ by $\textsf{F}$; the support of $\textsf{F}$ is a finite set $\cF$. The rate of the communication $\textsf{F}$ is defined as $\frac{1}{n}\log|\cF|$. Note that $\textbf{f}$, $\textsf{F}$ and $\cF$ implicitly depend on $n$.

\begin{definition}
\label{def:CR}
A \emph{common randomness (CR)} obtained from an interactive communication $\textsf{F}$ is a sequence of random variables $\textsf{J}^{(n)}$, $n\in\N$, which are functions of $X_{\mathcal{M}}^n$, such that for any $0<\epsilon<1$ and for all sufficiently large $n$, there exist $J_i=J_i(X_i^n,\textsf{F})$, $i = 1,2,\ldots,m$, satisfying $\text{Pr}[J_1=J_2=\cdots=J_m=\textsf{J}^{(n)}] \geq 1-\epsilon$.
\end{definition}

When $\textsf{J}^{(n)}=X_{\cM}^n$, we say that the terminals in $\cM$ have attained \emph{omniscience}. The communication $\textsf{F}$ which achieves this is called a \emph{communication for omniscience}. It was shown in Proposition~1 of \cite{CN04} that the minimum rate achievable by a communication for omniscience, denoted by $R_{\CO}$, is equal to $\displaystyle\min_{(R_1,R_2,\ldots,R_m)\in\mathcal{R}_{\CO}}\sum_{i=1}^mR_i$, where the region $\mathcal{R}_{\CO}$ is given by
\begin{equation}
\mathcal{R}_{\CO}=\biggl\{(R_i)_{i\in\cM}: \sum_{i\in B}R_i\geq \sum_{e\in\cE:e\subseteq B}\bw(e), B\subsetneq\cM\biggr\}.
\label{def:RCO}
\end{equation}
Henceforth, we will refer to $R_{\CO}$ as the ``minimum rate of communication for omniscience''. Note that the point $(R_1,R_2,\ldots,R_m)$ defined by $R_i=H(X_i)$ for all $i$ lies in $\mathcal{R}_{\CO}$, and hence $R_{\CO}\leq\sum_{i=1}^mH(X_i)<\infty$.

\begin{definition}
\label{def:SK}
A real number $R\geq 0$ is an \emph{achievable SK rate} if there exists a CR $\textsf{K}^{(n)}$, $n \in \N$, obtained from an interactive communication $\textsf{F}$ satisfying, for any $\epsilon > 0$ and for all sufficiently large $n$, $I(\textsf{K}^{(n)};\textsf{F})\leq \epsilon$ and $\frac{1}{n}H(\textsf{K}^{(n)}) \geq R-\epsilon$. The \emph{SK capacity} is defined to be the supremum among all achievable rates.  The CR $\textsf{K}^{(n)}$ is called a \emph{secret key (SK)}. 
\end{definition}

From now on, we will drop the superscript $(n)$ from both $\textsf{J}^{(n)}$ and $\textsf{K}^{(n)}$ to keep the notation simple. 

The SK capacity can be expressed as \cite[Theorem~1]{CN04}
\begin{equation}
\cC(\cM)=H(X_{\cM})-R_{\CO}.\label{omni}
\end{equation}
Other equivalent characterizations of $\cC(\cM)$ exist in the literature. One such characterization of SK capacity can be given via the notion of \emph{multivariate mutual information} defined as follows:
\begin{equation}
\Ixm\triangleq\min_{\cP}\Ipxm
\label{eq:I}
\end{equation}
with $\Ipxm\triangleq \frac{1}{|\cP|-1} \left[\sum_{A \in \cP} H(X_A) - H(X_{\cM}) \right]$ and the minimum being taken over all partitions $\cP=\{A_1,A_2,\cdots,A_{\ell}\}$ of $\cM$, of size $\ell \ge 2$. Note that $I(X_{\cM}^n)=n\Ixm$. The quantity $\Ixm$ is a generalization of the mutual information to a multiterminal setting; indeed, for $m=2$, we have $I(X_1,X_2)=I(X_1;X_2)$. It was shown in Theorem~1.1 of \cite{CZ10} and Theorem~4.1 of \cite{Chan14} that
\begin{equation}
\cC(\cM)=\Ixm.  \label{capeqmi}
\end{equation}
For the rest of this paper we shall use $\cC(\cM)$ and $\Ixm$ interchangeably. 

We will denote by $\cP^*$ the finest partition that achieves the minimum in \eqref{eq:I}. Theorem~5.2 of \cite{Chan14} guarantees that $\cP^*$ exists and is unique, and will henceforth be referred to as the \emph{fundamental partition}. In particular, we call the partition $\bigl\{\{1\},\{2\},\ldots,\{m\}\bigr\}$ consisting of $m$ singleton cells as the \emph{singleton partition} and denote it by $\cS$. The sources satisfying $\cP^*=\cS$ will be referred to as \emph{Type $\cS$ sources}.

We are now in a position to make the notion of communication complexity rigorous.

\begin{definition}
\label{def:RSKr}
A real number $R\geq 0$ is said to be an \emph{achievable rate of interactive communication for maximal-rate SK} if for all $\epsilon > 0$ and for all sufficiently large $n$, there exist \emph{(i)}~an interactive communication $\textsf{F}$ satisfying $\frac{1}{n}\log|\cF| \; \leq R+\epsilon$, and \emph{(ii)}~an SK $\textsf{K}$ obtained from $\textsf{F}$ such that $\frac{1}{n}H(\textsf{K})\geq \textbf{I}(X_{\mathcal{M}})-\epsilon$.

The infimum among all such achievable rates is called the \emph{communication complexity of achieving SK capacity}, denoted by $R_{\SK}$.
\end{definition}

The proof of Theorem~1 in \cite{CN04} shows that there exists an interactive communication $\textsf{F}$ that enables omniscience at all terminals and from which a maximal-rate SK can be obtained. Therefore, we have $R_{\SK}\leq R_{\CO}< \infty$. Hence, in terms of communication complexity, the sources that satisfy $R_{\SK}=R_{\CO}$ are the worst-case sources. We will henceforth refer to them as $R_{\SK}$-\emph{maximal} sources. Such sources do exist, as shown in Section~VI of \cite{MNY}.

\section{Upper Bound on $R_{\SK}$}\label{sec:ub}

In this section we derive an upper bound on $R_{\SK}$ based on the notion of \emph{decremental SK agreement} studied in \cite{Chan16}. The idea is to ``fractionally remove" hyperedges from the original hypergraph. To be precise, consider a hypergraphical source $X_\cM$ defined on the weighted hypergraph $\cH=(\cM,\bw)$. A non-negative vector $\bx$ satisfying $\bx\leq\bw$ (coordinatewise) is called a \emph{fractional packing} of the hypergraph $\cH$. For any fractional packing $\bx$, define a new hypergraphical source on the weighted hypergraph $\cH^\bx=(\cM,\bx)$. Observe that since $\bx\leq\bw$, the source defined on $\cH^\bx$ is obtained by ``removing" some randomness from $X_\cM$. 

We denote the relevant quantities for the source defined on $\cH^\bx$ by adding a superscript $^\bx$ to the original notation. For example, we use $X_\cM^\bx$, $R_{\CO}^\bx$, $R_{\SK}^\bx$ etc. It is easy to see that $I(X_\cM^\bx)\leq\Ixm$, since any SK generation protocol for $X_\cM^\bx$ is also a valid SK generation protocol for $X_\cM$.

To proceed, we need to introduce some notation. Define the set $\gub$ to be the set of all fractional packings $\bx$, satisfying the following constraints:
\begin{equation}
\bx\leq\bw, \label{gub:1}
\end{equation}
\begin{equation}
\exists \mathbf{r}=(r_i)_{i\in\cM} \text{ such that }\sum_{i\in B}r_i\geq \sum_{e\in\cE: e\subseteq B}\bx(e), \forall B\subsetneq\cM, \label{gub:2}
\end{equation}
\begin{equation}
\sum_{e\in\cE}\bx(e)-\sum_{i=1}^mr_i=\Ixm. \label{gub:3}
\end{equation}
Note that $\gub$ is non-empty since $\bw\in\gub$. This follows immediately by choosing $(r_1,r_2,\ldots,r_m)\in\mathcal{R}_{\CO}$ that achieves $R_{\CO}$ and by noting \eqref{omni} and the fact that $H(X_\cM)=\sum_{e\in\cE}\bw(e)$. Denote by $\gs$ the set of fractional packings $\bx$ satisfying $I(X_\cM^\bx)=\Ixm$, i.e., the fractional packing $\bx$ does not decrease the SK capacity. It is easy to see that $\bw\in\gs$, and hence it is non-empty.

We now state the upper bound to $R_{\SK}$ in the following theorem.

\begin{theorem}
\label{th:ub}
For a hypergraphical source model $X_\cM$ defined on the weighted hypergraph $\cH=(\cM,\bw)$ we have 
$$
R_{\SK}\leq\sum_{e\in\cE}\bx^*(e)-\Ixm, 
$$
where $\bx^*$ is an optimal solution for the linear program $\min\sum_{e\in\cE}\bx(e)$ subject to the constraints $\bx\in\gub$.
\end{theorem}

Since, $\bx^*\leq\bw$, by \eqref{omni} we have that the upper bound in Theorem~\ref{th:ub} is at least as good as $R_{\CO}$. We will need the following lemma in order to prove Theorem~\ref{th:ub}.

\begin{lemma}\label{lem:gub}
For a hypergraphical source $X_\cM$ defined on the weighted hypergraph $\cH=(\cM,\bw)$ we have 
$$
\gub=\gs.
$$
\end{lemma}

\begin{IEEEproof}
To begin with note that $H(X_\cM^\bx)=\sum_{e\in\cE}\bx(e)$. It is straightforward to see that the constraints in \eqref{gub:2} are nothing but the $\cR_{\CO}$ constraints of \eqref{def:RCO} for the source defined on $\cH^\bx$. Therefore, the constraint \eqref{gub:3} along with \eqref{omni} shows that a fractional packing $\bx\in\gub$ does not decrease the SK capacity, and so $\bx\in\gs$. Therefore, $\gub\subseteq\gs$.

On the other hand, any fractional packing $\bx\in\gs$ does not decrease the SK capacity. Hence, by \eqref{omni}, there exists a rate point $\mathbf{r}$ satisfying the $\cR_{\CO}$ constraints, i.e., the constraints in \eqref{gub:2}, as well as the constraint \eqref{gub:3}. So, $\bx\in\gub$ and hence $\gs\subseteq\gub$, which completes the proof.
\end{IEEEproof}

We are now in a position to prove Theorem~\ref{th:ub}.

\begin{IEEEproof}[Proof of Theorem~\ref{th:ub}]
To begin with, consider any $\bx\in\gub$. By Lemma~\ref{lem:gub}, we also have that $\bx\in\gs$. Since any SK generation protocol for $X_\cM^\bx$ is also a valid SK generation protocol for $X_\cM$, the fact that $I(X_\cM^\bx)=\Ixm$ implies that $R_{\SK}\leq R_{\SK}^\bx\leq R_{\CO}^\bx$. Using \eqref{omni}, we have $R_{\CO}^\bx=H(X_\cM^\bx)-I(X_\cM^\bx)=\sum_{e\in\cE}\bx(e)-\Ixm$. Therefore, combining the above results we have for any $\bx\in\gub$, $R_{\SK}\leq\sum_{e\in\cE}\bx(e)-\Ixm$. In order to get the best upper bound we simply choose $\bx^*$ which minimizes $\sum_{e\in\cE}\bx(e)$ among all possible $\bx\in\gub$.
\end{IEEEproof}

Before proceeding, we provide an example where we explicitly evaluate the upper bound in Theorem~\ref{th:ub}.

\begin{example}
\label{ex:square}
Consider the hypergraph $\cH=(\cM,\bw)$, with ${|\cM|}=4$ and weight vector $\bw$ given as follows:
\begin{align}
\bw(e)& = 1,  e=\{1,4\},\{2,3\},\{3,4\}\notag\\
           & = 2, e=\{1,2\}\notag\\
           & = 0, \text{ otherwise.}\notag
\end{align}
One can easily check that $\cP^*=\{\{1,2\},\{3\},\{4\}\}$, $\Ixm=1.5$ and $R_{\CO}=3.5$. Solving the linear program in Theorem~\ref{th:ub}, we see that the optimal fractional packing $\bx^*$ is given by $\bx^*(\{1,2\})=1.5$ and $\bx^*(e)=\bw(e)$ for all $e\neq\{1,2\}$. Thus, Theorem~\ref{th:ub} gives the upper bound $R_{\SK}\leq 3<R_{\CO}$.

It is not difficult to check that no integer-valued $\bx\neq\bw$ is possible without decreasing the SK capacity. Thus even if $\bw$ is integer-valued, removing a hyperedge by an integer amount or completely need not be optimal, and fractional removal is a better thing to do. 
\end{example}

We now turn our attention to evaluating the upper bound in Theorem~\ref{th:ub}. Evaluating the upper bound in Theorem~\ref{th:ub} requires the knowledge of $\Ixm$, which in turn can be calculated in strongly polynomial time as shown in \cite{Chan14}. Knowing $\Ixm$, the upper bound in Theorem~\ref{th:ub} can be computed in polynomial time. This is because the separation oracle for the constraints in \eqref{gub:2}, i.e., $\sum_{i\in B}r_i\geq \sum_{e\in\cE: e\subseteq B}\bx(e)=H(X^\bx_B|X^\bx_{B^c})$, for all $B\subsetneq\cM$, is but an instance of a submodular function minimization given by $\min_{B\subsetneq\cM}\biggl\{\sum_{i\in B}r_i-H(X^\bx_B|X^\bx_{B^c})\biggr\}\geq0$. The fact that submodular function minimization can be carried out in polynomial time (see Theorem~45.1 of \cite{Schrijver}), implies that the upper bound in Theorem~\ref{th:ub} can also be computed in polynomial time (see Theorem~5.10 of \cite{Schrijver}).

It turns out that for the special case of graphical models, i.e., when $\cE$ consists only of sets of size 2, the upper bound in Theorem~\ref{th:ub} reduces to a very simple expression.

\begin{theorem}\label{th:PIN}
For a source $X_\cM$ defined on a weighted graph $\cG=(\cM,\bw)$ we have 
$$
R_{\SK}\leq (m-2)\Ixm.
$$
\end{theorem}

Before proceeding, observe that for a graphical model, we have
\begin{equation}
\Ipxm=\frac{1}{|\cP|-1}\sum_{e\in\cE_\cP}\bw(e), \label{eq:delta}
\end{equation}
where $\cE_\cP$ denotes the set of edges $e\in\cE$ which are not contained in any parts of the partition $\cP$. We require the following lemma to prove Theorem~\ref{th:ub}.

\begin{lemma}\label{lem:PIN}
For a source $X_\cM$ defined on the graph $\cG=(\cM,\bw)$, we have $\gub=\gs=\{\bw\}$ only if $X_\cM$ is Type $\cS$.
\end{lemma}

\begin{IEEEproof}
We shall prove the lemma by contradiction. Suppose that $\gub=\{\bw\}$ but $\cP^*\neq\cS$. We will show that there exists a fractional packing $\bx\neq\bw$ which lies in $\gub$, thereby contradicting the assumption $\cP^*\neq\cS$. 

We first prove that for every $A\in\cP^*$ with $|A|\geq2$, there exists at least one edge of $\cG$ contained in $A$. Otherwise, any refinement of $\cP^*$ to some $\tilde{\cP}$ by arbitrarily splitting $A$ into two parts will satisfy $\cE_{\tilde{\cP}}=\cE_{\cP^*}$, and hence, $\displaystyle\sum_{e\in\cE_\cP}\bw(e)=\sum_{e\in\cE_{\tilde{\cP}}}\bw(e)$. That would imply $I_{\cP^*}(X_\cM)>I_{\tilde{\cP}}(X_\cM)$, violating the optimality of $\cP^*$ in \eqref{eq:I}. Hence, we can fix a $\tilde{e}\subseteq A$. 

Next we obtain a fractional packing $\bx\neq\bw$ by removing randomness from $\tilde{e}$. Let $\epsilon=\min_{\cP}(\Ipxm-I_{\cP^*}(X_\cM))$, the minimum being taken over all partitions $\cP\neq\cP^*$ which are not coarser versions of $\cP^*$. $\cP^*$ being the fundamental partition (and $\cP^*\neq\cS$) we have $\epsilon>0$; choose $0<\delta<\epsilon$. We claim that the fractional packing $\bx$, defined by $\bx(\tilde{e})=\bw(\tilde{e})-\delta$, and $\bx(e)=\bw(e)$, for all $e\neq\tilde{e}$, lies in $\gub$. This will violate the fact that $\gub=\{\bw\}$ and hence we will have the result by contradiction. To complete the proof we require to show that $\bx\in\gub(=\Gamma^*)$. 

To proceed, consider the graph $\cG^\bx$. Observe that by \eqref{eq:delta}, $I_{\cP^*}(X_\cM^\bx)=I_{\cP^*}(X_\cM)$ and hence $I_{\cP^*}(X_\cM^\bx)=\Ixm$. For any partition $\tilde{\cP}$ which is a coarser version of $\cP^*$, we have $I_{\tilde{\cP}}(X_\cM^\bx)=I_{\tilde{\cP}}(X_\cM)\geq\Ixm$, using \eqref{eq:delta}. On the other hand, consider any partition $\cP'$ which is not $\cP^*$ or a coarser version of it. By the choice of $\bx$ we have, $\displaystyle\sum_{e\in\cE_{\cP'}}\bx(e)\geq\sum_{e\in\cE_{\cP'}}\bw(e)-\delta$. Thus, using \eqref{eq:delta} we have, $I_{\cP'}(X_\cM^\bx)\geq\frac{1}{|\cP'|-1}\biggl[\sum_{e\in\cE_{\cP'}}\bw(e)-\delta\biggr]\geq I_{\cP'}(X_\cM)-\delta>I_{\cP'}(X_\cM)-\epsilon\geq\Ixm$. Hence, by \eqref{eq:I}, we have $I(X_\cM^\bx)\geq\Ixm$. Since $I(X_\cM^\bx)\leq\Ixm$ always holds, the result follows.
\end{IEEEproof}

We now prove Theorem~\ref{th:PIN}.

\begin{IEEEproof}[Proof of Theorem~\ref{th:PIN}]
We first show that the source $X_{\cM}^{\bx^*}$ defined on the weighted graph $\cG^{\bx^*}$ is Type $\cS$. If not, Lemma~\ref{lem:PIN} will imply that there exists a fractional packing $\bx'\leq\bx^*$ of $(\cM,\bx^*)$, satisfying $I(X_\cM^{\bx'})=I(X_\cM^{\bx^*})=\Ixm$. This in turn implies that $\bx'\in\gs=\gub$, thereby violating the optimality of $\bx^*$. Hence, $X_{\cM}^{\bx^*}$ is Type $\cS$. As a result we have, $\Ixm=I(X_\cM^{\bx^*})=I_{\cS}(X_\cM^{\bx^*})=\displaystyle\frac{1}{m-1}\sum_{e\in\cE}\bx^*(e)$. Therefore, the bound in Theorem~\ref{th:ub} reduces to $(m-2)\Ixm$ as required.
\end{IEEEproof}

We would like to remark here that for the special case of PIN models on graphs, which is obtained by restricting $\bw$ in the graphical model to be integer-valued, the same upper bound was also derived in Lemma~9 of \cite{MNY}, using protocols for SK generation developed in \cite{NN10} based on \emph{spanning tree packing}. 

It was shown in the proof of Theorem~\ref{th:PIN} that the source $X_{\cM}^{\bx^*}$ is Type $\cS$. Therefore, Theorem~6 of \cite{MNY} shows that $X_\cM^{\bx^*}$ is $R_{\SK}$-maximal. As a result, we have using Theorem~\ref{th:PIN} and \eqref{omni} that $R_{\SK}^{\bx*}=(m-2)\Ixm$. Since $X_{\cM}^{\bx^*}$ was obtained from $X_{\cM}$ by throwing away some randomness that did not affect the SK capacity, we believe that it should not affect $R_{\SK}$ as well. Hence, we conjecture that the upper bound in Theorem~\ref{th:PIN} is tight. In fact, this leads us to believe that the upper bound in Theorem~\ref{th:ub} is tight which is stated as a conjecture below.

\begin{conjecture*}
The upper bounds in Theorems~\ref{th:ub} and \ref{th:PIN} are tight.
\end{conjecture*}

\section{Lower Bounds on $R_{\SK}$}\label{sec:lb}

In this section, we restrict our attention to source models defined on graphs only and show that the lower bound derived in Theorem~2 of \cite{MNY} reduces to a very simple expression. For the hypergraphical model in its full generality computing that bound is difficult, except for the special case of Type $\cS$ sources on \emph{$t$-uniform hypergraphs} as shown in Theorem~6 of \cite{MNY}.

\begin{theorem}
For a source $X_{\cM}$ defined on a weighted graph $\cG=(\cM,\bw)$ we have
$$
R_{\SK}\geq\frac{|\cP^*|-2}{|\cP^*|-1}\sum_{e\in\cE_{\cP^*}}\bw(e),
$$
where $\cE_{\cP^*} \subseteq \cE$ denotes the set of edges not contained within any of the cells of the partition $\cP^*$.
\label{th:lb}
\end{theorem}

To prove this theorem we need to introduce some definitions and results from \cite{MNY}. We begin by introducing the definition of \emph{conditional multivariate mutual information}, which is a generalization of conditional mutual information to the multiterminal setting. The conditional multivariate mutual information of $X_{\cM}$ given a random variable $\BL$ is defined as\footnote{It should be noted that the definition of conditional multivariate mutual information used here is slightly different from what we called ``conditional multipartite information" in \cite{MNY}. However, the main results of all of these works continue to hold even with the current definition.} 
\begin{equation}
I(X_{\cM}|\BL)\triangleq\frac{1}{|\cP^*|-1}\biggl[\displaystyle\sum_{A\in\cP^*}H(X_A|\BL)-H(X_{\cM}|\BL)\biggr]. \label{cmi}
\end{equation}
The definition of $I(X_{\cM}|\BL)$ applies to any collection of jointly distributed random variables $X_{\cM}$; in particular it applies to the collection $X_{\cM}^n$. To be clear, 
$$
\Ixml\triangleq\displaystyle\frac{1}{|\cP^*|-1}\biggl[\sum_{A\in\cP^*}H(X_A^n|\BL)-H(X_{\cM}^n|\BL)\biggr].
$$

We use this definition of $\Ixml$ to extend to the multiterminal setting an asymptotic version of two-terminal Wyner common information (see \cite{Wyner75}) appearing in \cite{Tyagi13}.\footnote{One possible generalization of the non-asymptotic Wyner common information appearing in \cite{Wyner75} to the multiterminal setting is carried out in \cite{XLC13}.} 

\begin{definition}
\label{def:CI}
A \emph{(multiterminal) Wyner common information ($\CI_W$)} for $X_{\mathcal{M}}$ is a sequence of finite-valued functions $\textsf{L}^{(n)}=\textsf{L}^{(n)}(X_{\mathcal{M}}^n)$ such that $\frac{1}{n}I(X_{\mathcal{M}}^n|\textsf{L}^{(n)}) \to 0$ as $n \to \infty$. An \emph{interactive common information ($\CI$)} for $X_{\mathcal{M}}$ is a Wyner common information of the form $\textsf{L}^{(n)} = (\textsf{J},\textsf{F})$, where $\textsf{F}$ is an interactive communication and $\textsf{J}$ is a CR obtained from $\textsf{F}$. 
\end{definition}

Similar to Definitions~\ref{def:CR} and \ref{def:SK} we shall drop the superscript $(n)$ from $\textsf{L}^{(n)}$ for notational simplicity. Wyner common informations $\textsf{L}$ do exist: for example, the identity map $\textsf{L}=X_{\mathcal{M}}^n$ is a $\CI_W$. To see that $\CI$s $(\textsf{J},\textsf{F})$ also exist, observe that $\textsf{J}=X_{\mathcal{M}}^n$ and a communication $\textsf{F}$ enabling omniscience constitute a $\CI_W$, and hence, a $\CI$.

\begin{definition}
\label{def:CIrate}
A real number $R\geq 0$ is an \emph{achievable $\CI_W$ (resp.\ $\CI$) rate} if there exists a $\CI_W$ $\textsf{L}$ (resp.\ a $\CI$ $\textsf{L} = (\textsf{J},\textsf{F})$) such that for all $\epsilon > 0$, we have
$\frac{1}{n}H(\textsf{L})\leq R+\epsilon$ for all sufficiently large $n$. We denote the infimum among all achievable $\CI_W$ (resp.\ $\CI$) rates by $\CI_W(X_{\mathcal{M}})$ (resp. \ $\CI(X_{\mathcal{M}})$).
\end{definition}

With these definitions in hand, we summarize some of the results of \cite{MNY} needed for this section in the following theorem.
\begin{theorem}\label{th:old}
For a source $X_\cM$, we have
$$
\Ixm\leq\CI_W(X_\cM)\leq\CI(X_\cM)\leq H(X_\cM),
$$
and
$$
R_{\SK}\geq\CI(X_\cM)-\Ixm.
$$
\end{theorem}

To proceed, we will need a variant of the Lemma~7 of \cite{MNY} for graphs. 

\begin{lemma}
For any function $\BL$ of a source $X_{\cM}^n$ defined on a weighted graph $\cG=(\cM,\bw)$ with fundamental partition $\cP^*$, we have
$$\displaystyle\sum_{A\in\cP^*}I(X_A^n;\BL)\leq2H(\BL).
$$
\label{lem:mi}
\end{lemma}

The proof follows on the lines of Lemma~7 of \cite{MNY} and can be found in Appendix~\ref{app:proof} .The following lemma determines the minimum rate of interactive common information $\CI(X_{\cM})$ for graphical models.

\begin{lemma}
For the source $X_{\cM}$ defined on a weighted graph $\cG=(\cM,\bw)$, with fundamental partition $\cP^*$, we have
$$
\CI(X_{\cM})=\sum_{e\in\cE_{\cP^*}}\bw(e).
$$
\label{lem:CI}
\end{lemma}

\begin{IEEEproof}
To begin with, we observe that it suffices to prove that $CI_W(X_{\cM})=\sum_{e\in\cE_{\cP^*}}\bw(e)$. Indeed, consider $\BF$ to be a broadcast of all the random variables $\xi_e^n$ associated with the edges in $e\in\cE_{\cP^*}$. Let $\textsf{J}=(\xi_e^n:e\in\cE_{\cP^*})$. It is straightforward to verify that $I(X_\cM^n|\BJ,\BF)=0$, since $\sum_{A\in\cP^*}H(X_A^n|\BJ,\BF)=H(X_{\cM}^n|\BJ,\BF)=\sum_{e\notin\cE_{\cP^*}}\bw(e)$. Therefore, the pair $(\BJ,\BF)$ constitutes a $\CI$ whose rate is $\sum_{e\in\cE_{\cP^*}}\bw(e)$. Hence, we would have $\CI(X_\cM)\leq\sum_{e\in\cE_{\cP^*}}\bw(e)=\CI_W(X_{\cM})$, which along with Theorem~\ref{th:old} would give the result.

The proof of $CI_W(X_{\cM})=\sum_{e\in\cE_{\cP^*}}\bw(e)$ follows along the lines of the proof of Theorem~6 of \cite{MNY}. At first choosing $\BL=(\xi_e^n:e\in\cE_{\cP^*})$, it follows immediately that $\Ixml=0$. Thus, $\BL$ is a $CI_W$ for $X_{\cM}^n$, and so $\CI_W(X_\cM)\leq\sum_{e\in\cE_{\cP^*}}\bw(e)$. Next, we prove $CI_W(X_{\cM})\geq\sum_{e\in\cE_{\cP^*}}\bw(e)$. To proceed, let $\BL$ be any function of $X_{\cM}^n$. Then, we have 
\begin{align}
I(X_{\mathcal{M}}^n | \textsf{L}) 
   & = \frac{1}{|\cP^*|-1}\biggl[\sum_{A\in\cP^*}H(X_A^n|\BL)-H(X_{\cM}^n|\BL)\biggr]\notag\\
   & = \frac{1}{|\cP^*|-1}\biggl[\sum_{A\in\cP^*}\left(H(X_A^n,\BL)-H(\BL)\right)\notag\\
   &\hspace{2cm}-H(X_{\cM}^n,\BL)+H(\BL)\biggr]\notag\\
   & = H(X_{\cM}^n,\BL)-H(\BL)\notag\\
   &\hspace{0.5cm}+\frac{1}{|\cP^*|-1}\biggl[\sum_{A\in\cP^*}\left(H(X_A^n,\BL)-H(X_{\cM}^n,\BL)\right)\biggr]\notag\\
   & = H(X_{\cM}^n)-H(\BL)-\frac{1}{|\cP^*|-1}\sum_{A\in\cP^*}H(X_{A^c}^n|X_A^n,\BL)\label{cmi0}\\
   & = H(X_{\mathcal{M}}^n) - H(\textsf{L}) \notag\\
   &\hspace{0.5cm}- \frac{1}{|\cP^*|-1} \sum_{A\in\cP^*}\biggl[H(X_{\cM}^n) - H(X_A^n) - H(\BL|X_A^n)\biggr] \label{cmi1} \\
   & = n\sum_{e\in\cE}\bw(e)\left(1-\frac{|\cP^*|}{|\cP^*|-1}\right)\notag\\
   &\hspace{0.5cm}+\frac{n}{|\cP^*|-1}\biggl[\sum_{e\in\cE}\bw(e)+\sum_{e\in\cE_{\cP^*}}\bw(e)\biggr]- H(\textsf{L}) \notag\\
   &\hspace{0.5cm}+ \frac{1}{|\cP^*|-1} \sum_{A\in\cP^*} H(\textsf{L}|X_A^n) \label{cmi2}\\
   & = \frac{n}{|\cP^*|-1}\sum_{e\in\cE_{\cP^*}}\bw(e)\notag\\
   &\hspace{0.5cm} - \frac{1}{|\cP^*|-1} \left[ \sum_{A\in\cP^*} I(X_A^n;\BL)-H(\BL)\right] \notag \\
   & = \frac{n}{|\cP^*|-1}\left(\sum_{e\in\cE_{\cP^*}}\bw(e)-\frac{1}{n}H(\BL)\right)\notag\\
   &\hspace{0.5cm}-\frac{1}{|\cP^*|-1} \left[ \sum_{A\in\cP^*} I(X_A^n;\BL)-2H(\BL)\right] \notag \\
   & \geq \frac{n}{|\cP^*|-1}\left(\sum_{e\in\cE_{\cP^*}}\bw(e)-\frac{1}{n}H(\BL)\right), \label{cmi3}
\end{align}
where \eqref{cmi0} and \eqref{cmi1} follow from the fact that $\BL$ is a function of $X_{\cM}^n$; \eqref{cmi2} follows from the fact that $H(X_{\cM}^n)=n\sum_{e\in\cE}\bw(e)$ and $\sum_{A\in\cP^*}H(X_A^n)=n\biggl[\sum_{e\in\cE}\bw(e)+\sum_{e\in\cE_{\cP^*}}\bw(e)\biggr]$; and \eqref{cmi3} is due to Lemma~\ref{lem:mi}.

Now, consider $\textsf{L}$ to be any $\CI_W$ so that for any $\epsilon > 0$, we have $\frac{1}{n}I(X_{\mathcal{M}}^n | \textsf{L}) < \frac{\epsilon}{(|\cP^*|-1)}$ for all sufficiently large $n$. The bound in \eqref{cmi3} thus yields $\frac{1}{n}H(\BL)> \sum_{e\in\cE_{\cP^*}}\bw(e)-\epsilon$ for all sufficiently large $n$. Hence, it follows that $\CI_W(X_{\cM}) \ge \sum_{e\in\cE_{\cP^*}}\bw(e)$. Therefore, we obtain $CI_W(X_{\cM})=\sum_{e\in\cE_{\cP^*}}\bw(e)$.
\end{IEEEproof} 

We now prove Theorem~\ref{th:lb}.

\begin{IEEEproof}[Proof of Theorem~\ref{th:lb}]
For the source $X_{\cM}$ we have
\begin{align}
R_{\SK} & \geq \CI(X_{\cM})-\Ixm \label{th:lb:1}\\
              &  = \sum_{e\in\cE_{\cP^*}}\bw(e)-\frac{1}{|\cP^*|-1}\sum_{e\in\cE_{\cP^*}}\bw(e) \label{th:lb:2}\\
              &  = \frac{|\cP^*|-2}{|\cP^*|-1}\sum_{e\in\cE_{\cP^*}}\bw(e)\notag
\end{align}
where, \eqref{th:lb:1} follows from Theorem~\ref{th:old} and \eqref{th:lb:2} follows from Lemma~\ref{lem:CI}.
\end{IEEEproof}

Unfortunately, it turns out that the lower bound in Theorem~\ref{th:lb} is not tight in general as illustrated by the following example.

\begin{example}
\label{ex:loose}
Consider the source $X_\cM$ defined on a weighted graph $\cG=(\cM,\bw)$ with ${|\cM|}=4$. The weight vector $\bw$ is given by $\bw(e)=1$ for $e=\{1,2\},\{1,3\},\{2,3\}$ and $\{3,4\}$, and $\bw(e)=0$ otherwise. Thus, $\cE=\{\{1,2\},\{1,3\},\{2,3\},\{3,4\}\}$. It is straightforward to verify that $\cP^*=\{\{1,2,3\},\{4\}\}$ and $\Ixm=1$. Theorem~\ref{th:lb} gives the lower bound $R_{\SK}\geq 0$ for this source. However, it is clear that the combined observations of terminal 1 and 2, i.e., $X_{\{1,2\}}^n$, is completely independent of $X_4^n$. Hence, a communication of positive rate would certainly required for achieving SK capacity. Thus, $R_{\SK}>0$, which implies that the lower bound in Theorem~\ref{th:lb} is loose.
\end{example}

\section{Concluding Remarks}\label{sec:conc}

The upper bound in Theorem~\ref{th:ub} is the first reported upper bound on $R_{\SK}$ for any instance of the multiterminal source model of \cite{CN04}. We showed that this bound is at least as good as the obvious upper bound of $R_{\CO}$, and can in fact be stronger as illustrated by Example~\ref{ex:square}. We further show that this upper bound can be computed in polynomial time. We believe that this upper bound is tight. Due to the lack of a proof we have left it as a conjecture. We have also evaluated the lower bound on $R_{\SK}$ stated in \cite[Theorem~2]{MNY} for the special case of graphical source models. The evaluation enabled us to construct an example showing that the lower bound is not tight in general.

\section*{Acknowledgements}

The authors would like to thank Navid Nouri for stimulating discussions which helped in writing this paper and gain a better understanding of the problem at hand.

\begin{appendices}

\section{Proof of Lemma \ref{lem:mi}}\label{app:proof}

The lemma essentially follows from Lemma~7 in \cite{MNY}. A weighted hypergraph $(\cM,\bw)$ is \emph{$t$-uniform} if all hyperedges in $\cE = \{e \in 2^{\cM}: \bw(e) > 0\}$ have size exactly $t$. In particular, a weighted graph satisfies this definition with $t = 2$. We then have the following lemma.

\begin{lemma}[\cite{MNY}, Lemma~7] Let $X_{\cM}$ be a source defined on a $t$-uniform weighted hypergraph. For any $n \in \N$ and any function $\BL$ of $X_{\cM}^n$, we have
$$
\sum_{i = 1}^m I(X_i^n;\BL) \le t\,H(\BL).
$$
\label{lem:appendix}
\end{lemma}

It should be clarified that Lemma~7 in \cite{MNY} is stated only for hypergraphs with integer-valued weight functions $\bw$. However, this restriction is not essential for the proof given in \cite{MNY}, so that it applies to any real-valued weight function $\bw$ just as well. We will need the above lemma only for the case of graphical source models (i.e., $t = 2$).

To prove our Lemma~\ref{lem:mi}, consider the given weighted graph $\cG = (\cM,\bw)$, with edge set $\cE = \{e \in 2^{\cM} : \bw(e) > 0\}$, and the fundamental partition $\cP^* = (A_1,\ldots,A_k)$ of the corresponding source $X_{\cM}$.  From $X_{\cM}$, we construct a closely related graphical source $\tX_{\cV}$ on a vertex set $\cV$, as described next. 

For each pair of cells $A_i,A_j$ of $\cP^*$, with $i < j$, let $\cE_{i,j}$ denote the set of edges of $\cG$ with one endpoint in $A_i$ and the other in $A_j$. We further let $\cE_i$ be the set of edges of $\cG$ that are contained within $A_i$, $i = 1,2,\ldots,k$. Now, let $\{a_1,\ldots,a_k\}$ and $\{a_1',\ldots,a_k'\}$ be two disjoint sets of size $k = {|\cP^*|}$ each, and let $\cV = \{a_1,\ldots,a_k\} \cup \{a_1',\ldots,a_k'\}$. Define a weight function $\tbw$ on $2$-subsets of $\cV$ as follows: for each pair of integers $i < j$, we set $\tbw(\{a_i,a_j\}) = \sum_{e \in \cE_{i,j}} \bw(e)$; and for $i = 1,2,\ldots,k$, set $\tbw(\{a_i,a_i'\}) = \sum_{e \in A_i} \tbw(e)$. For all other $2$-subsets $\tilde{e}$ of $\cV$, we set $\tbw(\tilde{e}) = 0$.

We take ${\tX}_{\cV}$ to be a source defined on the weighted graph $(\cV,\tbw)$. To be precise, let $\xi_e$, $e \in \cE$, be the random variables associated with the edges of the graphical source $X_{\cM}$. In $\tX_{\cV}$, we associate with each $2$-subset (edge) $\tilde{e}$ of $\cV$, a random variable $\tilde{\xi}_{\tilde{e}}$ as below:
$$
\tilde{\xi}_{\tilde{e}} = 
\begin{cases}
(\xi_e: e \in \cE_{i,j}) & \text{ if } \tilde{e} = \{a_i,a_j\} \\
(\xi_e: e \in \cE_i) & \text{ if } \tilde{e} = \{a_i,a_i'\} \\
0 & \text{ otherwise.}
\end{cases}
$$
As usual, for any $v \in \cV$, $\tX_{v}$ refers to the random variable $(\xi_{\tilde{e}}: v \in \tilde{e})$. Observe, in particular, that $\tX_{a_i} = X_{A_i}$, for $i = 1,2,\ldots,k$.

Now, to complete the proof of Lemma~\ref{lem:mi}, we note that for any function $\BL$ of $X_{\cM}^n$, we have
\begin{align*}
\sum_{A\in\cP^*}I(X_A^n;\BL)  \ \ &= \ \ \sum_{i=1}^k I(\tX_{a_i}^n; \BL) \\
&\le \ \ \sum_{v \in \cV} I(\tX_{v}^n; \BL)  \\
&\le \ \ 2H(\BL)
\end{align*}
by Lemma~\ref{lem:appendix} (with $t = 2$).

\end{appendices}


\begin{thebibliography}{99}

\bibitem{CN04}
I.~Csisz{\'a}r and P.~Narayan, ``Secrecy capacities for multiple terminals,'' \emph{IEEE Trans.\ Inf.\ Theory}, vol.\ 50, pp.\ 3047--3061, Dec.\ 2004.

\bibitem{Tyagi13}
H.~Tyagi, ``Common information and secret key capacity," \emph{IEEE Trans.\ Inf.\ Theory}, vol.\ 59, no.\ 9, pp.\ 5627--5640, Sep.\ 2013.

\bibitem{MNY}
M.~Mukherjee, N.~Kashyap, Y.~Sankarasubramaniam, ``On the public communication needed to achieve SK capacity in the multiterminal source model," Arxiv:1507.02874.

\bibitem{Yao79}
A. C.~Yao, ``Some complexity questions related to distributed computing,'' in \emph{Proc.\ 11th Annu.\ ACM Symp.\ Theory of Computing (STOC)}, 1979.

\bibitem{CZ10}
C.~Chan and L.~Zheng, ``Mutual dependence for secret key agreement," in \emph{Proc.\ 44th Annual Conference on Information Sciences and Systems (CISS)}, 2010.

\bibitem{ESS10}
S.\ El Rouayheb, A.\ Sprintson, and P.\ Sadeghi, ``On coding for cooperative data exchange,'' in 
\emph{Proc.\ 2010 IEEE Inf.\ Theory Workshop (ITW 2010)}, Cairo, Egypt, 6--8  Jan.\ 2010, pp.\ 1--5.

\bibitem{CW14}
T. A.~Courtade and R. D.~Wesel, ``Coded cooperative data exchange in multihop networks,"  \emph{IEEE Trans.\ Inf.\ Theory}, vol.\ 60, no.\ 2, pp.\ 1136--1158, Feb.\ 2014.

\bibitem{CHISIT14}
T. A.~Courtade and T. R.~Halford, ``Coded cooperative data exchange for a secret key," in \emph{Proc.\ 2014 IEEE Int.\ Symp.\ Inf.\ Theory (ISIT 2014)}, Honolulu, Hawai'i, USA, June 29 -- July 4, 2014, pp.\ 776--780.

\bibitem{CH14}
T. A.~Courtade and T. R.~Halford, ``Coded cooperative data exchange for a secret key," Arxiv:1407.0333v1.

\bibitem{NYBNR10}
S.~Nitinawarat, C.~Ye, A.~Barg, P.~Narayan and A.~Reznik, ``Secret key generation for a pairwise independent network model,'' \emph{IEEE Trans.\ Inf.\ Theory}, vol.\ 56, pp.\ 6482--6489, Dec.\ 2010.

\bibitem{NN10}
S.~Nitinawarat and P.~Narayan, ``Perfect omniscience, perfect secrecy and Steiner tree packing,'' \emph{IEEE Trans.\ Inf.\ Theory}, vol.\ 56, no.\ 12, pp.\ 6490--6500, Dec.\ 2010. 

\bibitem{Chan16}
C.~Chan, A.~Al-Bashabsheh and Q.~Zhou, ``Incremental and decremental secret key agreement", submitted to \emph{2016 IEEE Int.\ Symp.\ Inf.\ Theory (ISIT 2016)}, Barcelona, Spain, July 10 -- 15, 2016.

\bibitem{Chan14}
C.~Chan, A.~Al-Bashabsheh, J.~Ebrahimi, T.~Kaced and T.~Liu, ``Multivariate mutual information inspired by secret key agreement," in \emph{Proc.\ of IEEE}, vol.\ 103, no.\ 10, pp.\ 1883-1913, Oct.\ 2015.


\bibitem{Schrijver}
A.~Schrijver, \emph{Combinatorial Optimization: Polyhedra and Efficiency}, Volume A--C, Springer, 2004.

\bibitem{Wyner75} 
A. D.~Wyner, ``The common information of two dependent random variables,'' \emph{IEEE Trans.\ Inf.\ Theory}, vol.\ IT-21, no.\ 2, pp.\ 163--179, Mar.\ 1975.

\bibitem{XLC13}
G.~Xu, W.~Liu and B.~Chen, ``Wyner's common information: Generalizations and a new lossy source coding interpretation," Arxiv:1301.2237v1.

\end{thebibliography}
\end{document}